**Fast Object Removal Attacks on Safety-Critical Video-based Perception Systems**

**Mohammad Imtiaz Hasan***
Glenn Department of Civil Engineering
Clemson University, Clemson, SC 29634
ORCID: https://orcid.org/0009-0006-9799-6773
Email: hasan2@clemson.edu

**M Sabbir Salek, Ph.D**
Senior Engineer
National Center for Transportation Cybersecurity and Resiliency
Greenville, SC 29607
Email: msalek@clemson.edu

**Nathan Jones**
School of Computing
Clemson University, Clemson, SC 29634
Email: najones@clemson.edu

**Mashrur Chowdhury, Ph.D, P.E.**
Eugene Douglas Mays Chair of Transportation
Glenn Department of Civil Engineering
Clemson University, Clemson, SC 29634
Email: mac@clemson.edu

**Rong Ge, Ph.D**
Professor
School of Computing
Clemson University, Clemson, SC 29634
Email: rge@clemson.edu

*Corresponding Author

## ABSTRACT

**Objectives:** By leveraging data from video-based perception systems, intelligent transportation systems (ITS) support safety-critical applications that improve road safety. However, adversaries may manipulate video frames to compromise downstream perception modules, causing failures in safety-critical functions and increasing risks to vulnerable road users. This paper aims to develop and evaluate a near-real-time object-removal attack model where a cyber-attacker removes targeted objects from video while preserving visual consistency in the reconstructed frames.

**Methods**: The end-to-end attack pipeline consists of four stages: localizing targets in each frame, retrieving coherent patches from earlier frames, blending them using context-aware alpha compositing, and reconstructing attacked frames.

**Findings**: Experiments at an intersection on the South Carolina Connected Vehicle Testbed (SC-CVT) show that reconstructed frames have high global similarity to the originals, with frame-level Peak Signal to Noise Ratio (PSNR) above 40 dB and Structural Similarity Index Measure (SSIM) above 0.996. Using the YOLO-based detector, the attack reduces object detections by up to 97.59% and achieves a frame-level attack success rate of 94.48%. Across the evaluated detectors and frame resolutions, the mean execution time ranges from 0.074 to 0.172 seconds per frame on GPU hardware, indicating near-real-time performance in testing. The forensic evaluation using several pretrained tamper-detection models shows limited ability to distinguish reconstructed from authentic frames.

**Novelty**: This paper presents a novel attack model and an end-to-end framework for near-real-time targeted object removal attack on a video-based safety-critical system. Prior studies have not demonstrated this type of attack on video-based systems in near real time.

**Practical Applications:** The findings suggest that video-based perception is vulnerable to stealthy object removal attacks that can degrade the performance of safety-critical applications by reducing object detectability. These findings can help develop mitigation strategies against adversarial object removal attacks that threaten safety-critical applications, such as vision-based pedestrian safety systems.

## INTRODUCTION

Intelligent transportation systems (ITS) have the potential to improve road safety and traffic efficiency by leveraging infrastructure-augmented perception systems consisting of sensors, edge computing devices and communication technologies (Yu et al. 2022). Roadside infrastructure-mounted video cameras are among the widely used sensing modalities in ITS environments. Image streams acquired by the infrastructure-mounted mono or multi-camera systems are processed by deep learning-based models, such as convolutional neural networks, running on roadside edge computing devices for object detection and classification (Jebamikyous and Kashef 2022). Recent developments in these models have significantly improved object detection accuracy, at the same time maintaining real-time inference capabilities (Chandrashekhar et al. 2024). The output from these models is subsequently utilized to generate real-time safety messages (Islam et al. 2020; Enan et al. 2024). These safety messages are leveraged by ITS to enable safety-critical applications such as pedestrian presence warning, collision warning, queue warning, and emergency management. Accurate and reliable object detection is therefore a critical requirement within the ITS architecture.

Despite the advancement in vision-based perception, numerous studies have demonstrated that deep neural networks remain vulnerable to adversarial attacks (Moosavi-Dezfooli et al. 2016; Rahman et al. 2025). An adversarial attack generally introduces carefully crafted perturbations that cause a deep learning-based object detector to misclassify an object or reduce its detection confidence. Perturbations can be physical, such as printed perturbations and camouflage patterns (Eykholt et al. 2017), or digital, such as optimization-based image perturbations (Goodfellow et al. 2014). Although these attacks often achieve high attack success rates, the target object usually remains visually present in the scene and detectable to the detectors. Consequently, both human observers and alternative sensing modalities may still recognize the target even when it is misclassified.

A less explored yet potentially more severe threat is semantic object removal attacks (Wang et al. 2025). Instead of perturbing image pixels to change the detector output, object-removal attacks alter the image stream by replacing the target object with plausible background content. This replacement may be generated through video inpainting or obtained from spatially corresponding regions in temporally adjacent frames (Habeeb and Manikandan 2019). From the perspective of the perception system, the pedestrian or vehicle effectively disappears entirely from the frames while the surrounding portion of the removed object remains visually consistent. Consequently, the subsequent object detection models and human observers will not detect any object in the frames. Such attacks are particularly concerning for infrastructure-augmented perception because widely deployed stationary edge computing devices are vulnerable to unauthorized adversarial access (Xiao et al. 2019). Furthermore, erroneous object detection information caused by the attack may lead to the failure of safety-critical applications in ITS.

Although several LiDAR-spoofing-based object removal attacks have been reported in the literature (Sato et al. 2025; Suzuki et al. 2025), no research has examined object removal attacks against widely deployed video-based perception in ITS environment, to the best of our knowledge. Furthermore, despite advancements in deep-learning-based image inpainting models, inpainting remains a time-consuming task, executing object removal attacks using video inpainting in real time remains challenging (Li et al. 2022).

To this end, this study presents a near-real-time semantic object removal attack model against safety-critical video-based perception systems shown in **Figure 1**. In the first stage, the attack detects and tracks target objects using a real-time object detector and generates corresponding bounding boxes. For each bounding box region in a frame, a search is performed over previously captured frames to identify the most recent spatially coherent background for the target region. The selected patch from the temporally adjacent frame is then used to replace the target object region in the current frame using context-aware alpha blending. A loss-based seam minimization is then applied to further reduce the visual inconsistencies around the modified region. Unlike conventional adversarial perturbation attacks, this approach removes the object from the input frames of the detector and manipulates the entire detector response rather than altering the classifier decision boundaries. As the attacked region is synthesized from spatially consistent historical frames, the manipulated frame demonstrates fewer visual evidences than conventional image editing approaches, making attack detection challenging for available tamper-detection tools. In contrast to attacks that primarily target LiDAR-based perception, our attack approach focuses on widely deployed infrastructure-mounted camera-based systems and therefore examines a more near-term and practically relevant attack surface in ITS environments. Moreover, the near-real-time execution of the framework in this paper demonstrates its potential for application to live video streams, establishing a more realistic threat model for safety-critical perception systems. To the best of our knowledge, no prior studies have demonstrated near-real-time or real-time object removal attacks on video frames, highlighting the practical significance of this work.

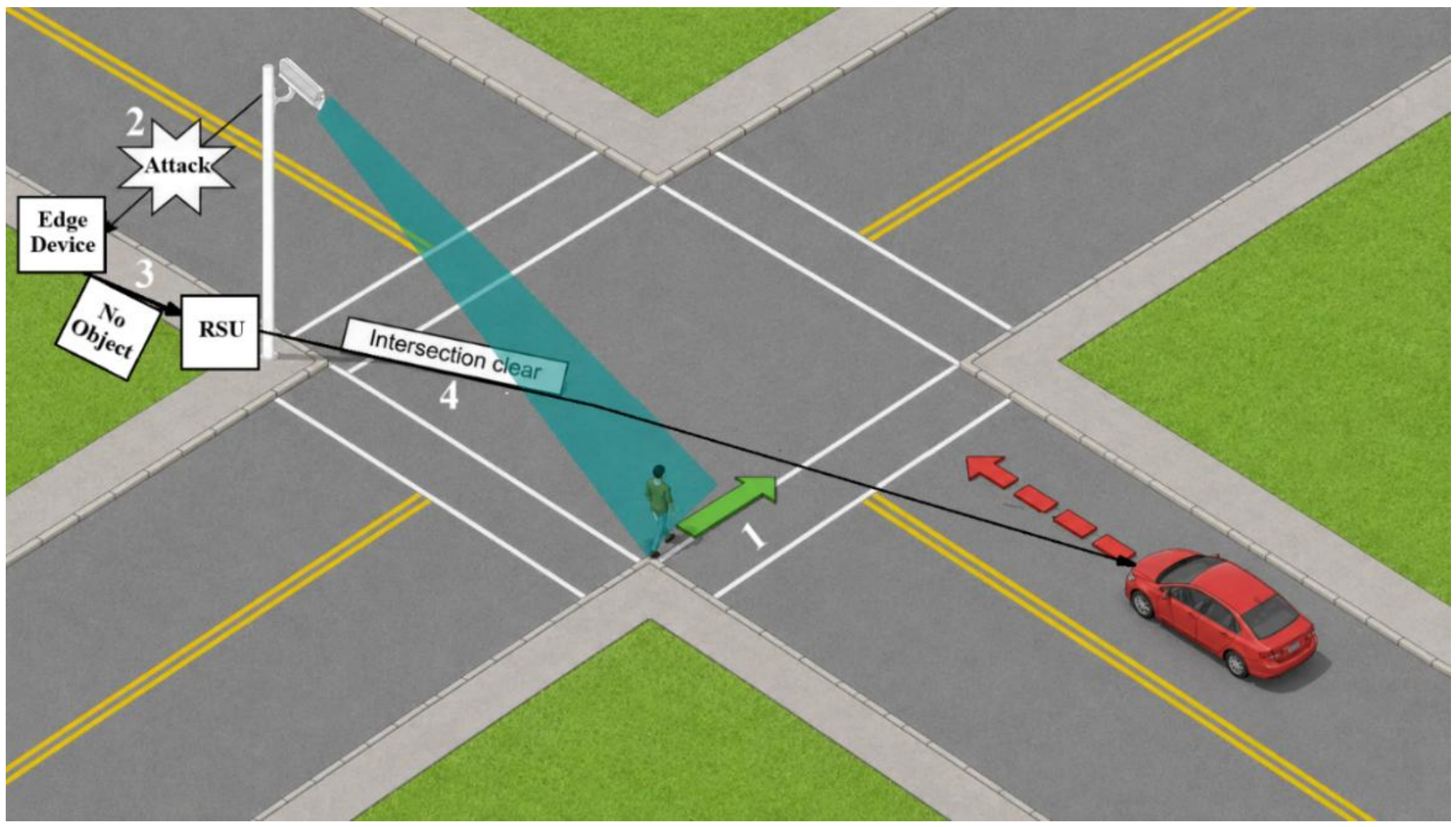

**Figure 1 Effect of targeted object removal attack on ITS safety-critical functions showing 1) pedestrian entering intersection, 2) attack on the video feed, 3) edge device sending no object detection to RSU, and 4) RSU sends 'intersection clear' message to vehicles**

To validate the attack under realistic operating conditions, we utilize field data collected from the South Carolina Connected Vehicle Testbed (SC-CVT), an operational connected vehicle testbed (CVT) equipped with infrastructure-mounted cameras and vehicular communication technologies. The SC-CVT captures real-world interactions among vehicles, pedestrians, and cyclists under varying environmental conditions, thereby providing a realistic evaluation environment for infrastructure-augmented perception. This enables a real-world evaluation of our object removal attack in terms of attack success rate, reconstructed video quality, and computational latency under practical conditions.

The main contributions of this work are as follows:

- A novel attack model designed to induce a targeted object removal attack in a video-based perception system. The attack model integrates target localization, historical patch retrieval to replace the target object, context-aware alpha blending to reduce boundary mismatch, and a seam-loss-based smoothing scheme to minimize visual discrepancies.
- An attack execution framework capable of executing the attack in near real time, in terms of safety-critical applications, while maintaining visual consistency of the image frames, and
- An evaluation of the attack model on data collected from real-world CVT across two frame resolutions commonly used in transportation video-based perception systems.

## LITERATURE REVIEW

In this section, we discuss prior research on object removal approaches for generating tampered videos, as well as studies that employ object removal attacks against different perception systems. The discussion highlights key methods and limitations in these areas to contextualize and motivate the attack presented in this work.

There are two techniques researchers have employed for object removal from image frames- inpainting and copying temporally adjacent frames with spatial coherence (Habeeb and Manikandan 2019). Inpainting is a process to fill the blank space in an image frame with a visually plausible background. In recent years, deep learning-based inpainting has become a popular research area within computer vision, targeting use cases such as cultural relic restoration, watermark removal, film and television production, and digital forensics (Quan et al. 2024; Kim et al. 2019). Most of the deep learning-based inpainting studies can be categorized into four major approaches: 3D CNN, shift-based approach, flow-guided propagation and attention-based approach (Quan et al. 2024). 3D CNNs combine temporal restoration with image inpainting to address the temporal dimension (Chang et al. 2019; Hu et al. 2020). Shift-based approach enables a 2D CNN to learn temporal information from video without resorting to a computationally expensive 3D CNN (Ke et al. 2021; Zou et al. 2021). A flow-guided approach fills the blank spaces in the frame by propagating pixels from boundary regions (Gao et al. 2020; Yu et al. 2019). Several inpainting studies have leveraged attention modules to adequately capture temporal context (Li et al. 2020; Woo et al. 2019). Although these approaches produce visually consistent outputs, they are computationally expensive for high-resolution image frames and not suitable for real-time image inpainting during an attack. Temporal copy-paste-based object removal copies pixels from a specific area, also known as the patch area, from nearby frames or the same frame, to mask the target object (Singh and Aggarwal 2021). While temporal copy-paste-based techniques are simpler and faster, automatically selecting an appropriate patch remains challenging because of spatial and temporal variations in shadows and illumination. Furthermore, additional actions such as resizing or rotation are often required to generate visually coherent output (Singhal and Gandhani 2015).

Attack against the vision-based perception system is a well-studied area. However, most of these attacks are adversarial attacks aiming to generate misclassification by detection modules (Moosavi-Dezfooli et al. 2017, 2016; Goodfellow et al. 2014; Eykholt et al. 2017; Rahman et al. 2025). Existing studies on object removal attacks against the perception systems mainly focus on spoofing LiDAR (Suzuki et al. 2025; Hau et al. 2021; Sato et al. 2025). None of these attacks target the widely deployed video-based perception module, and the impact such attacks have on the safety-critical aspect of ITS. To address these gaps, we present a novel attack model and develop an end-to-end pipeline to execute a near-real-time object removal attack against an infrastructure-mounted camera-based perception system.

## METHODS

This section presents the attack model for the near-real-time targeted object-removal attack and introduces its mathematical formulation. The first subsection defines the adversarial objective and describes the attacker's knowledge and capabilities. The second subsection provides an overview of the attack framework and explains its four main stages.

### Attack Model

The threat we consider in this work is an object removal attack against a video-based perception system, in which an adversary manipulates image frames to prevent a physically present object from being detected by downstream object detection models. The attack is performed by altering the pixel content inside target object regions and replacing it with visually plausible background-consistent content. Consequently, the attack reduces detector efficacy while preserving the scene's natural appearance. The adversarial objective is therefore not only to corrupt the image, but to suppress object detection without introducing evidence that would reveal a visual cue of tampering.

#### *Attack Objective*

Let $I_t \in \mathbb{R}^{H \times W \times 3}$denote the clean input frame at time $t$, where $H$ and $W$ are the image height and width, respectively, and let $\mathcal{F}(\cdot)$represent the object detection module that outputs a set of detections. For a clean frame, **Equation 1** defines the detector output:

$$\mathcal{Y}_t = \mathcal{F}(I_t) = \{y_t^{(k)}\}_{k=1}^{K_t}, \tag{1}$$

where each detection $y_t^{(k)}$ includes at least a spatial region and an object confidence score, $K_t$ is the number of detections produced for the frame. The adversary generates a manipulated frame $\hat{I}_t$ such that the number of valid detections in $\hat{I}_t$is reduced relative to $I_t$, eliminating one or more target objects entirely. The attack objective can be expressed as minimizing the detector's object detection response over a target set of objects $\mathcal{T}_t \subseteq \mathcal{Y}_t$ while maintaining visual plausibility given by **Equation 2**:

$$\min_{\hat{I}_t} \sum_{y \in \mathcal{T}_t} S(y \mid \hat{I}_t) \tag{2}$$

subject to $\hat{I}_t \approx I_t$ outside the attacked regions, where $S(y \mid \hat{I}_t)$ denotes the detection confidence score associated with the target object $y$ in the manipulated frame. In other words, the attack seeks to suppress target detections while preserving scene consistency in the non-target regions.

*Adversary Knowledge and Capability*

The attack model assumes that the adversary has unauthorized access to an edge-computing device, the video stream or image frame sequence before it reaches the perception module. Under this assumption, the adversary can modify the incoming visual data prior to object detection and downstream perception. This threat model is motivated by the widespread deployment of static edge devices and the expanded attack surface they introduce, which increase their exposure to unauthorized access (Xiao et al. 2019). The adversary does not require prior knowledge of the spatial location of the target object within the frame. Instead, the attack pipeline incorporates an object detection model that automatically identifies candidate target regions in each frame. The detected bounding boxes are then used to execute the subsequent attack stages. Because the attack pipeline only requires bounding-box outputs, it is not tied to any specific detector architecture. Any object detector capable of identifying the target class and returning spatial bounding boxes can be integrated if it is fast enough to enable real-time attacks.

The attack is intentionally constrained in two important ways. First, modifications are restricted to spatial regions automatically identified by the integrated object detector, thereby avoiding trivial full-frame corruption. Second, the reconstructed regions must remain visually coherent with the surrounding background so that the manipulation does not introduce visual anomalies that are easily detectable by human inspection or downstream forensic checks. These constraints make the attack more stealthy, visually realistic and more challenging to detect than unconstrained image degradation.

**Overview of the Attack Pipeline**

The attack method addresses object removal in image sequences by combining four stages, as shown in **Figure 2**: 1) target-region localization, 2) temporal clean-patch retrieval, 3) context-aware alpha blending in an expanded region, and 4) frame reconstruction over all target boxes. Together, these stages identify target regions of interest in each frame, search for historical clean patches, replace the pixel values of the target region with the pixel values from the patch, blend the inserted content into the surrounding context to minimize boundary discontinuities, and reconstruct the frame. This design is appropriate for object removal attack because it models the practical scenario in which an attacked object region is removed and substituted with scene content estimated from earlier temporal observations.

Let $\{I_t\}_{t=1}^{T}$ denote a sequence of RGB frames, where $I_t \in \mathbb{R}^{H \times W \times 3}$ is the frame at time $t$. For each frame $I_t$, an object localization function produces a set of bounding boxes, which are defined in **Equation 3**:

$$\mathcal{B}_t = \{b_t^{(k)}\}_{k=1}^{K_t} \tag{3}$$

where $K_t$ is the number of detected objects, and each bounding box is given by **Equation 4**:

$$b_t^{(k)} = (x_{1,t}^{(k)}, y_{1,t}^{(k)}, x_{2,t}^{(k)}, y_{2,t}^{(k)}) \tag{4}$$

The objective of the attack is to construct a modified frame $\hat{I}_t$ in which each target region is replaced by a temporally valid background-consistent patch while preserving local visual coherence. Below, we detail the four stages of this attack pipeline.

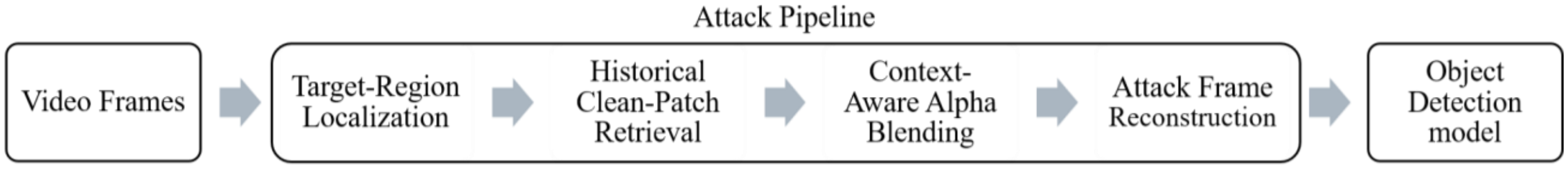

**Figure 2: Object removal attack stages**

*Target-Region Localization*
The first step is to determine the bounding boxes of target objects to be removed. A generic object localization function is applied independently to each frame to estimate the object set $\mathcal{B}_t$, defined in **Equation 3**, where each box defines the coordinates of a region to be processed. The method does not depend on a specific detector family. Hence, any detector capable of returning rectangular bounding boxes can be used, as long as it keeps the overall attack time within the time required to perform the attack in real-time.

For a given frame $I_t$, the localization operator can be expressed as per **Equation 5**:

$$\mathcal{F}(I_t) \rightarrow \mathcal{B}_t. \quad (5)$$

Each box $b_t^{(k)}$ defines a target region, $\Omega_t^{(k)}$ in **Equation 6**:

$$\Omega_t^{(k)} = \{(x, y) \mid x_{1,t}^{(k)} \leq x < x_{2,t}^{(k)}, y_{1,t}^{(k)} \leq y < y_{2,t}^{(k)}\}, \quad (6)$$

which corresponds to the spatial area where object content is to be replaced. The method assumes that the localization stage provides sufficiently accurate bounding boxes for downstream restoration, but it does not otherwise constrain the inference mechanism. When multiple target objects are present in the same frame, each region is processed individually. This yields a set of candidate removal regions whose order of processing defines the sequence of local modifications that eventually produce the final attacked frame.

*Temporal Clean-Patch Retrieval*
Once a target region $\Omega_t^{(k)}$ has been identified, the method seeks a temporally earlier observation at the same pixel coordinates that was free of the object. The rationale is that the immediately earlier frames may preserve the background appearance of the target region in terms of lighting, color, shadow, etc., before the object entered the scene or before the attacked region became visually occupied. By retrieving a clean patch from an earlier frame of the same scene, the method reuses previously observed background information rather than synthesizing the missing content with an image inpainting model. However, the retrieval strategy presented in this work assumes limited viewpoint variation within the temporal search window, which is consistent with fixed infrastructure-augmented camera-based perception systems. Under this assumption, previously observed image regions provide valid spatial references for replacing the corresponding target locations.

For a target box $b_t^{(k)}$, the method searches backward through prior frames $I_{t-1}, I_{t-2}, \ldots, I_1$ and checks whether the current box overlaps any detected object region in the candidate reference frame. Let $b_t^{(k)}$ be the current target box and $\mathcal{B}_s$ be the set of boxes in a previous frame $I_s$, where $s < t$. A previous frame is considered valid if **Equation 7** holds:

$$\forall b_s^{(j)} \in \mathcal{B}_s, \Omega_t^{(k)} \cap \Omega_s^{(j)} = \emptyset. \quad (7)$$

where $\Omega_s^{(j)}$ represents the pixel region of the $j$-th detected bounding box in $I_s$. The overlap test is implemented using rectangle intersection, so a frame is rejected if any prior detected box intersects the current target region. This condition ensures that the extracted reference patch is not occupied by object pixels from the earlier frame. The selected source index for the object $k$ in frame $t$ is therefore given by **Equation 8**:

$$s^\star = \max\{s \in \{1, \dots, t-1\} \mid \forall b_s^{(j)} \in \mathcal{B}_s,\ \Omega_t^{(k)} \cap \Omega_s^{(j)} = \emptyset\}, \tag{8}$$

i.e., the most recent previous frame satisfying the non-overlap condition. The clean patch extracted is given by **Equation 9**:

$$P_t^{(k)} = I_{s^\star}[y_{1,t}^{(k)}\!:\!y_{2,t}^{(k)},\ x_{1,t}^{(k)}\!:\!x_{2,t}^{(k)}]. \tag{9}$$

This formulation imposes temporal consistency by favoring the nearest object-free historical appearance of the same spatial region. If no prior frame exists, the region is left unchanged, thereby avoiding the introduction of arbitrary or unsupported content into the reconstruction.

Because the source patch and target region must have identical spatial dimensions, the retrieved patch is resized if the resolution changes between frames. Let the target width and height be given by **Equation 10**:

$$w_t^{(k)} = x_{2,t}^{(k)} - x_{1,t}^{(k)}, h_t^{(k)} = y_{2,t}^{(k)} - y_{1,t}^{(k)}. \tag{10}$$

The resized patch is then defined by **Equation 11**:

$$\tilde{P}_t^{(k)} = \mathcal{R}(P_t^{(k)}; h_t^{(k)}, w_t^{(k)}), \tag{11}$$

where $\mathcal{R}(\cdot)$denotes bilinear interpolation. Bilinear resizing is used to preserve spatial smoothness and avoid unwanted visual distortions when the dimensions differ between the patch and the target region. This method acts as a safeguard for a successful attack even when the frame resolution changes mid-stream.

*Context-Aware Alpha Blending*

Naive patch insertion often produces sharp visual seams because the inserted patch and the current-frame surroundings may differ in illumination, texture, or local perspective. To address this issue, the method blends the inserted patch with the local context inside an expanded region of interest (ROI), while preserving the interior of the target patch. This strategy is adapted from (Porter and Duff 1984) and enables the object area to be fully replaced while smoothing the transition in the immediate surrounding border.

For each target box $b_t^{(k)}$, an expanded ROI is defined by a padding parameter $p > 0$, given by **Equation 12**:

$$\tilde{b}_t^{(k)}(p) = (\tilde{x}_1, \tilde{y}_1, \tilde{x}_2, \tilde{y}_2), \tag{12}$$

with $\tilde{x}_1 = \max(0, x_1 - p)$, $\tilde{y}_1 = \max(0, y_1 - p)$, $\tilde{x}_2 = \min(W, x_2 + p)$, $\tilde{y}_2 = \min(H, y_2 + p)$, where $(x_1, y_1, x_2, y_2)$ abbreviates $b_t^{(k)}$ for readability. The expanded ROI includes the original target area and a surrounding context border to ensure a smooth transition. Within the expanded ROI, a patch canvas is created by placing the resized patch $\tilde{P}_t^{(k)}$ at the target location. Let $C_p$ denote this patch canvas and $B_p$ denote the background content cropped from the current frame and overlaid on the expanded ROI. The final blended result is controlled by an alpha mask $A_p(u, v)$, defined over each pixel $(u, v)$ in the expanded ROI. The alpha mask is constructed to satisfy two conditions: 1) full pixel preservation inside the target patch, and 2) smooth decay from the patch boundary into the outer border. Let $d(u, v)$ denote the Euclidean distance from the pixel $(u, v)$ to the nearest point on the target patch rectangle. Then the ring alpha before smoothing is given by **Equation 13**:

$$\alpha_{\text{ring}}(u, v) = 1 - \min\left(1, \frac{d(u, v)}{p}\right). \tag{13}$$

To obtain a smoother transition profile, the method applies a cubic smoothstep function defined in **Equation 14**:

$$\phi(x) = x^2(3 - 2x), x \in [0,1]. \tag{14}$$

Hence, the final alpha mask is given in **Equation 15**:

$$A_p(u,v) = \begin{cases} 1, & (u,v) \in \Omega_t^{(k)}, \\ \phi(\alpha_{\text{ring}}(u,v)), & (u,v) \in \widetilde{\Omega}_t^{(k)}(p) \setminus \Omega_t^{(k)}, \end{cases} \tag{15}$$

where $\widetilde{\Omega}_t^{(k)}(p)$ denotes the expanded ROI corresponding to $\tilde{b}_t^{(k)}(p)$. The interior of the patch, therefore, remains unchanged, while the surrounding ring decays continuously toward the original frame content. The blended ROI is computed by **Equation 16**:

$$R_t^{(k)}(p) = A_p \odot C_p + (1 - A_p) \odot B_p, \tag{16}$$

where all operations are pointwise over the expanded ROI. This expression can be interpreted as a soft compositing rule in which the inserted patch dominates near the patch center, while the original frame dominates farther away from the boundary. As a result, the method reduces visible discontinuities without altering the intended replacement inside the target region.

The extent of contextual blending strongly affects the quality of the attacked frame. If the padding width $p$ is too small, the transition from the inserted patch to the background may remain visually abrupt; if $p$ is too large, excessive blending may dilute structural consistency and introduce unnecessary distortion into the surrounding region. To address this trade-off, the method performs discrete seam-loss minimization over a set of candidate padding widths and selects the one that minimizes a seam mismatch criterion.

Let $\mathcal{P} = \{p_1, p_2, \dots, p_M\}$ denote the candidate set of padding values. For each $p \in \mathcal{P}$, the method computes a blended ROI $R_t^{(k)}(p)$ and evaluates a seam-loss on narrow bands across the boundary separating the preserved patch interior and the surrounding blended context. This loss evaluates the boundary mismatch between the blended region and its surrounding background. Consider the top, bottom, left, and right boundary bands around the target patch inside the expanded ROI. Let $\Gamma_m^{\text{in}}$ and $\Gamma_m^{\text{out}}$ denote the paired interior and exterior band regions for the side $m \in$ {top,bottom,left,right}, and let $|\Gamma_m|$ denote the number of pixels in the band. The seam-loss for the candidate $p$ is defined as per **Equation 17**:

$$\mathcal{L}_{\text{seam}}(p) = \frac{1}{M_p} \sum_{m=1}^{M_p} \frac{1}{|\Gamma_m|} \sum_{(u,v)\in\Gamma_m} \rho(R_t^{(k)}(p)_{\text{in}}(u,v) - R_t^{(k)}(p)_{\text{out}}(u,v)), \tag{17}$$

where $\rho(.)$ is a general loss function and $M_p$ is the number of valid boundary sides available for the current geometry. This loss penalizes intensity discontinuity across the patch boundary and therefore directly measures how perceptually smooth the insertion is likely to be.

The optimal padding width is obtained by **Equation 18**:

$$p^\star = \arg\min_{p\in\mathcal{P}} \mathcal{L}_{\text{seam}}(p). \tag{18}$$

The selected ROI can be defined as **Equation 19**:

$$R_t^{(k)\star} = R_t^{(k)}(p^\star). \tag{19}$$

This optimization procedure allows the blending radius to adapt to scene structure and local appearance rather than being fixed globally.

*Frame Reconstruction*

After the optimal blended ROI has been computed for each target region, the frame is reconstructed by sequentially updating the current image with the selected local restorations. Let $\hat{I}_t^{(0)} = I_t$ denote the initialization of the restored frame. For each detected target region $k = 1,2,\dots,K_t$, the method retrieves a valid historical patch, computes the optimal context-aware blend, and replaces the corresponding expanded ROI in the current reconstructed frame. This produces sequential updates given by **Equation 20**:

$$\hat{I}_t^{(k)} = \Psi(\hat{I}_t^{(k-1)}, b_t^{(k)}, \tilde{P}_t^{(k)}, p^\star), \tag{20}$$

where $\Psi(\cdot)$denotes the local ROI replacement operator using the optimized blended result. The final restored frame is therefore defined as **Equation 21**:

$$\hat{I}_t = \hat{I}_t^{(K_t)}. \tag{21}$$

More explicitly, if the expanded ROI associated with object $k$ is $\widetilde{\Omega}_t^{(k)}(p^\star)$, then:

$$\hat{I}_t(x,y) = \begin{cases} R_t^{(k)\star}(x,y), & (x,y) \in \widetilde{\Omega}_t^{(k)}(p^\star), \\ \hat{I}_t^{(k-1)}(x,y), & \text{otherwise.} \end{cases} \tag{22}$$

This operation is repeated for all target regions in the frame. If no valid historical clean patch is found for a particular object, no update is applied to that region, thereby preserving reconstruction reliability and preventing detectable modifications. The resulting frame reconstruction mechanism is therefore grounded in two principles: temporal consistency through historical patch reuse, spatial coherence through context-aware alpha blending and seam-loss minimization. Together, they form a detector-agnostic and model-flexible method for generating visually plausible attack frames.

## EXPERIMENTAL SETUP AND EVALUATION

This section presents the experimental setup and evaluation criteria for the near-real-time object-removal attack. The first subsection describes the dataset and preprocessing steps, while the second outlines the experimental settings and parameter configurations. The third subsection introduces the four sets of evaluation metrics used in this study.

### Dataset and Preprocessing

The experiments were conducted on video data collected from SC-CVT, a real-world CVT located in Clemson, South Carolina. The source dataset consists of 28 videos of $1920 \times 1080$ resolution, from which frames were sampled at 0.5 frames/s for frame-level evaluation, generating 191 frames. Each frame was subsequently downsampled to $1280 \times 720$ pixels using the bilinear interpolation method, resulting in a lower-resolution dataset. These two resolutions were selected to represent high-definition and moderate-resolution traffic-surveillance settings, commonly found in the literature (Zhang et al. 2022). The evaluation, therefore, used two temporally aligned frame sets: 191 original 1080p frames and 191 corresponding 720p frames. This design enabled a comparison of attack effectiveness and reconstruction quality across two commonly used resolutions while preserving scene content and temporal order. Preprocessing was limited to chronological organization and resolution conversion.

### Object Detector and Attack Implementation

We selected two representative object detectors for target localization and bounding-box generation: a YOLO-based detector, YOLO26 and a Faster R-CNN-based detector (Jocher et al. 2026; Ren et al. 2016). The YOLO configuration used the pretrained yolo26x.pt model with a confidence threshold of 0.20 and an inference image size of 352 pixels. The Faster R-CNN configuration used fasterrcnn_resnet50_fpn_v2 with pretrained COCO weights and a detection-score threshold of 0.20. Comparing these models enabled us to examine whether attack behavior varies between one-stage and two-stage detection architectures. Consistent misdetection across both detector families would provide evidence that the attack is not limited to a single detection architecture.

During the experiment, object detection was restricted to the COCO person category, effectively making our experiments a pedestrian removal attack. We used the same input frame sets, target class, restoration pipeline, and evaluation protocol for both detectors, while detector-specific preprocessing and inference procedures followed their respective frameworks. All experiments were executed on a NVIDIA A100 GPU using CUDA acceleration. The YOLO detector used GPU-based half-precision inference, whereas Faster R-CNN was executed with its default single-precision floating-point (FP32) configuration.

The influence of the seam-loss formulation was also investigated by comparing an $L_1$-based boundary discrepancy with a quadratic $L_2$-type discrepancy during context-aware blending. For each detector and resolution, the same detected bounding boxes were used when comparing the two seam-loss formulations. The resulting reconstructions were evaluated using seam-loss, patch-region PSNR, patch-region SSIM, attack success rate, and processing time.

**Evaluation Method**

The attack was evaluated from four complementary perspectives: image similarity and reconstruction consistency, attack effectiveness, computational efficiency, and tamper detectability by external pretrained forensic models. The following subsections present the details of these perspectives and define the evaluation metrics.

*Image Similarity and Reconstruction Consistency*

Image similarity and reconstruction consistency were assessed using both frame-level and patch-level image similarity measures. At the frame level, Peak Signal-to-Noise Ratio (PSNR) and Structural Similarity Index Measure (SSIM) were computed between the original and reconstructed frames to quantify the overall visual change introduced by the attack. Because the original target region contains the removed object, these metrics reflect similarity to the original frame. PSNR captures pixel-wise distortion while SSIM reflects perceptual similarity in terms of local structure, luminance, and contrast.

In addition to full-frame evaluation, localized image similarity was measured within the restored object regions. Patch PSNR and patch SSIM were computed within the manipulated bounding-box regions to quantify the magnitude and structural characteristics of localized modifications. This distinction is important because frame-level scores may be dominated by unmodified image regions, whereas patch-level metrics isolate the areas directly affected by the attack.

PSNR is given in **Equation 23** (Zou et al. 2021):

$$\mathrm{PSNR} = 10\log_{10}\left(\frac{L^2}{\mathrm{MSE}}\right), \tag{23}$$

where the MSE is the Mean Squared Error given in **Equation 24**:

$$\mathrm{MSE} = \frac{1}{N}\sum_{i=1}^{N}(I_i - R_i)^2, \tag{24}$$

$I_i$ and $R_i$ are corresponding pixel values in the original and reconstructed images, $N$ is the number of evaluated pixels, and $L$ is the maximum possible pixel value. For 8-bit images, $L = 255$.

SSIM is defined in **Equation 25** (Kim et al. 2019):

$$\mathrm{SSIM}(I, R) = \frac{(2\mu_I\mu_R + C_1)(2\sigma_{IR} + C_2)}{(\mu_I^2 + \mu_R^2 + C_1)(\sigma_I^2 + \sigma_R^2 + C_2)}, \tag{25}$$

where $\mu_I$ and $\mu_R$ are the local means, $\sigma_I^2$ and $\sigma_R^2$ are the local variances, $\sigma_{IR}$ is the local covariance, and $C_1$ and $C_2$ are stabilization constants.

The same formulations were applied to the full frame and to the manipulated bounding-box regions to compute $\mathrm{PSNR}_{\mathrm{frame}}$, $\mathrm{SSIM}_{\mathrm{frame}}$ and $\mathrm{PSNR}_{\mathrm{patch}}$, $\mathrm{SSIM}_{\mathrm{patch}}$, respectively.

*Attack Effectiveness*

Attack effectiveness was measured using the reduction in object detections after reconstruction. For each frame, the number of detections in the original frame was compared with the number of detections in the corresponding reconstructed frame. The attack success rate (ASR) was then defined at the frame level as the proportion of attacked frames in which detections of all pedestrians were suppressed, which is given in **Equation 26**:

$$\mathrm{ASR}_{\mathrm{frame}} = \frac{N_s}{N_A} \tag{26}$$

where $N_s$ represents the number of frames with all target detection missed by the downstream object detector and $N_A$ is the total number of attacked frames.

This experiment evaluates whether the removal attack successfully suppresses pedestrian detection while preserving visual plausibility. Frames with no original detections were excluded from the attack success computation, since suppression cannot be assessed when no target object is initially detected. Additionally, frames for which no historical clean patch was available were also excluded from the computation. In general, this can occur for the first few frames and as the object progresses spatially through the frames, previously unoccupied spatial regions become occupied, increasing the likelihood that a suitable object-free historical patch is available.

*Runtime Performance*

To assess practical applicability, the runtime per frame was measured for both the pedestrian detection and reconstruction stages. Detection time was recorded as the total inference time required to load a frame and localize pedestrians across the frames of a sequence, while reconstruction time was recorded as the total time required to retrieve clean reference patches, perform blending, and generate the reconstructed frames. Runtime was reported as the mean processing time per frame. According to the SAE J2735 standard, the maximum allowable latency for safety-critical applications is 100 milliseconds (V2X Core Technical Committee, n.d.). Latency above this threshold may disrupt safety-critical functions. Therefore, we aimed to keep the attack execution time as low as possible so that downstream safety functions could continue operating without revealing the attack through system failure. In this study, a runtime between 100 and 200 milliseconds is considered near real time during evaluation.

*Tamper Detectability*

To determine whether existing manipulation-detection systems can reliably distinguish original frames from reconstructed frames generated by the pedestrian removal attack, three pretrained tamper-detection models- DeCLIP, MVSS-net plus and PSCC-net were evaluated on the original and reconstructed frames (Dong et al. 2023; Smeu et al. 2025; Liu et al. 2022). These models were selected because they were developed for image manipulation detection or localization, including splicing and copy-move scenarios that share characteristics with the patch-replacement process of our attack.

Each pretrained model was used in an inference-only setting to produce a frame-level manipulation confidence score. These confidence scores were then used for binary classification of the original and reconstructed frames. Original frames were assigned the ground-truth label of authentic, while reconstructed frames were assigned the ground-truth label of manipulated or tampered. Performance was reported using accuracy, F1-score, and the area under the receiver operating characteristic curve (ROC-AUC). The decision threshold was determined by evaluating the model confidence scores over the observed confidence score range and selecting the threshold that maximized the F1-score. This evaluation was designed to assess the extent to which publicly available pretrained tamper-detection models can distinguish reconstructed from authentic frames.

## RESULTS AND DISCUSSION

In this section, we present the experimental results and discuss their implications. The reconstructed set of frames used in the experiments is represented by the naming convention- <object detector model>_<frame resolution>_<seam-loss>.

*Image Similarity and Reconstruction Consistency*

The image-similarity analysis revealed a clear distinction between full-frame and patch-level metrics, consistent with the localized nature of the attack. As shown in **Figure 3**, the YOLO-based configuration achieved a frame-level PSNR of approximately 40.18 dB and a SSIM value of 0.9980 across all resolutions and seam-loss configurations, while the patch-level PSNR values remained around 15.6 dB for all configurations and SSIM values ranging from 0.4973 to 0.5502. These frame-level values indicate very high global similarity between the reconstructed and original frames, as PSNR values above 40 dB and SSIM values close to 1.0 generally correspond to limited overall image distortion. In contrast, the lower patch-level PSNR and SSIM values indicate substantial localized changes within the attacked regions. However, these patch-level values should not be interpreted as poor reconstruction quality because the target region in the original frames contains the target object, whereas the reconstructed patch contains replacement background content. Instead, they confirm that the attack introduces strong semantic modification locally while preserving the overall visual appearance of the frame.

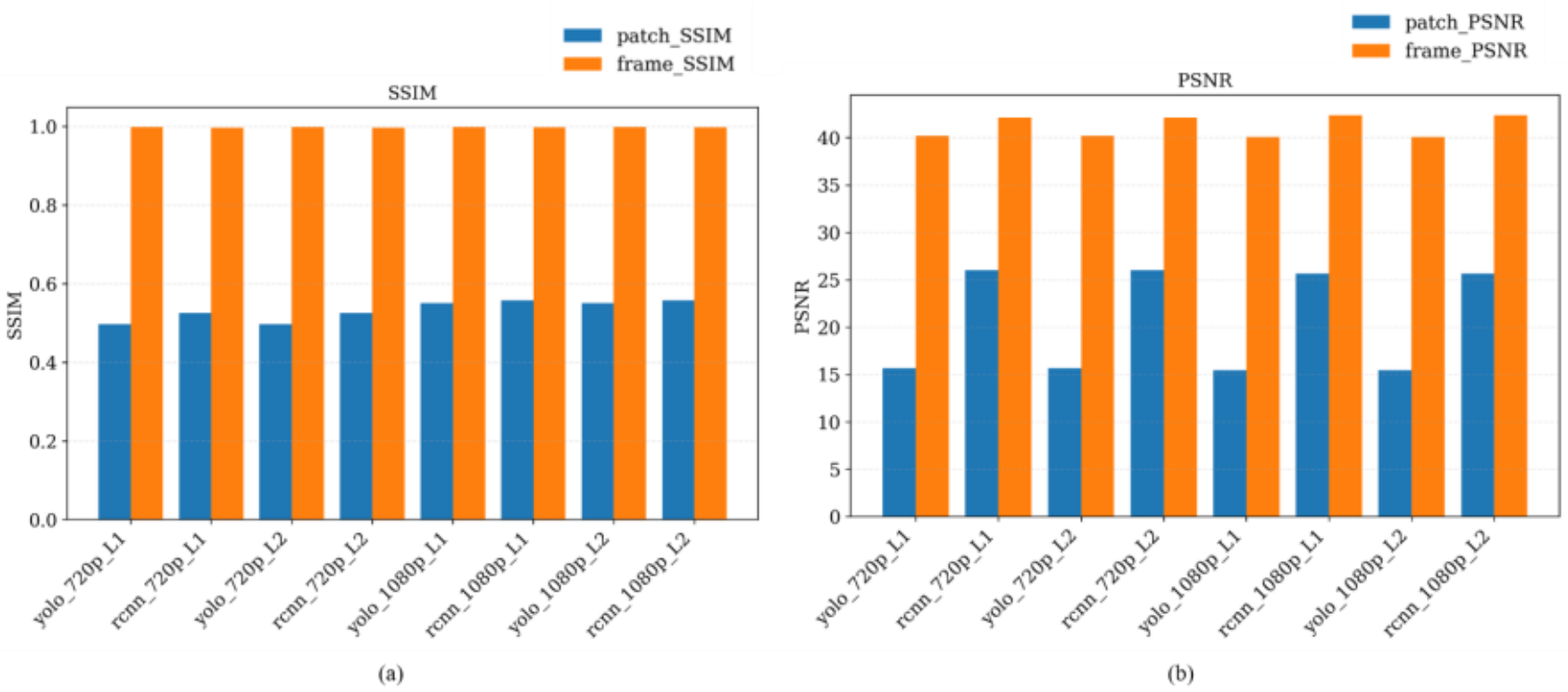

**Figure 3 (a) SSIM and (b) PSNR scores for different configurations**

Similar to the YOLO-based configuration results, the Faster R-CNN-based configurations produced high frame-level scores with PSNR and SSIM values above 42.13 dB and 0.997, respectively and substantially lower patch-level scores, with PSNR values between 26.04 and 25.62 dB and SSIM values between 0.5250 and 0.5579 across all configurations. These results indicate that our attack can generate visually consistent attack frames while introducing strong semantic modification at the patch-level, irrespective of object detection model, resolution and loss configurations. Although frame-level and patch-level score comparison follow similar behavior for both object detectors, patch-level PSNR scores for the Faster R-CNN-based configurations were higher than the YOLO-based configurations. Our experimental analysis indicated that Faster R-CNN generated a large number of false-positive pedestrian detections. Consequently, these regions underwent minor semantic change after restoration, producing higher patch-level PSNR values. A qualitative representation of the original and reconstructed frames is presented in **Figure 4** for both the YOLO and Faster R-CNN-based models.

*Attack Effectiveness*

The attack-effectiveness results in **Figure 5** indicate substantial suppression of target-object detections after reconstruction. Under the YOLO-based configuration, the object detection reduction rate reached 97.59% for $1280 \times 720$ resolution and 91.21% for $1920 \times 1080$ resolution. The corresponding frame-level attack success rates were 94.48% and 88.96%, respectively. The same object detection reduction rate and frame-level ASR values were obtained under both the $L_1$ and quadratic seam-loss formulations. This reduction in attack effectiveness can be attributed to two factors. First, the detector identified a larger number of pedestrians in the higher-resolution frames, and suitable clean historical patches were not available for some of these targets. Consequently, those pedestrians could not be fully reconstructed and remained detectable in the attacked frames. Second, inference variability caused a small number of targets that were not detected in the original frames to be detected in the reconstructed frames, even though their corresponding spatial regions were not modified. In these cases, minor variations in the detector output may have caused the confidence scores to cross the fixed detection threshold.

The Faster R-CNN-based configuration also exhibited substantial detection suppression, although the measured values were lower than those obtained with YOLO at both resolutions. The object detection reduction rates were approximately 87.07%, while the corresponding frame-level ASR values ranged between 74.69% and 70.55% across all configurations. Our analysis found that the Faster R-CNN-based detection model produced a lot of false positive pedestrian detection, which were replaced with the same background and consequently detected again in the reconstructed image. Although the object detection reduction rate and ASR score were low for the Faster R-CNN-based detector, these findings indicate that the attack remained effective across the two evaluated detector architectures. Furthermore, the near-identical ASR values obtained under the $L_1$ and quadratic seam-loss formulations further indicate that the choice of boundary loss had limited influence on detector suppression under the tested conditions. This suggests that the attack effectiveness was driven primarily by successful target-region replacement.

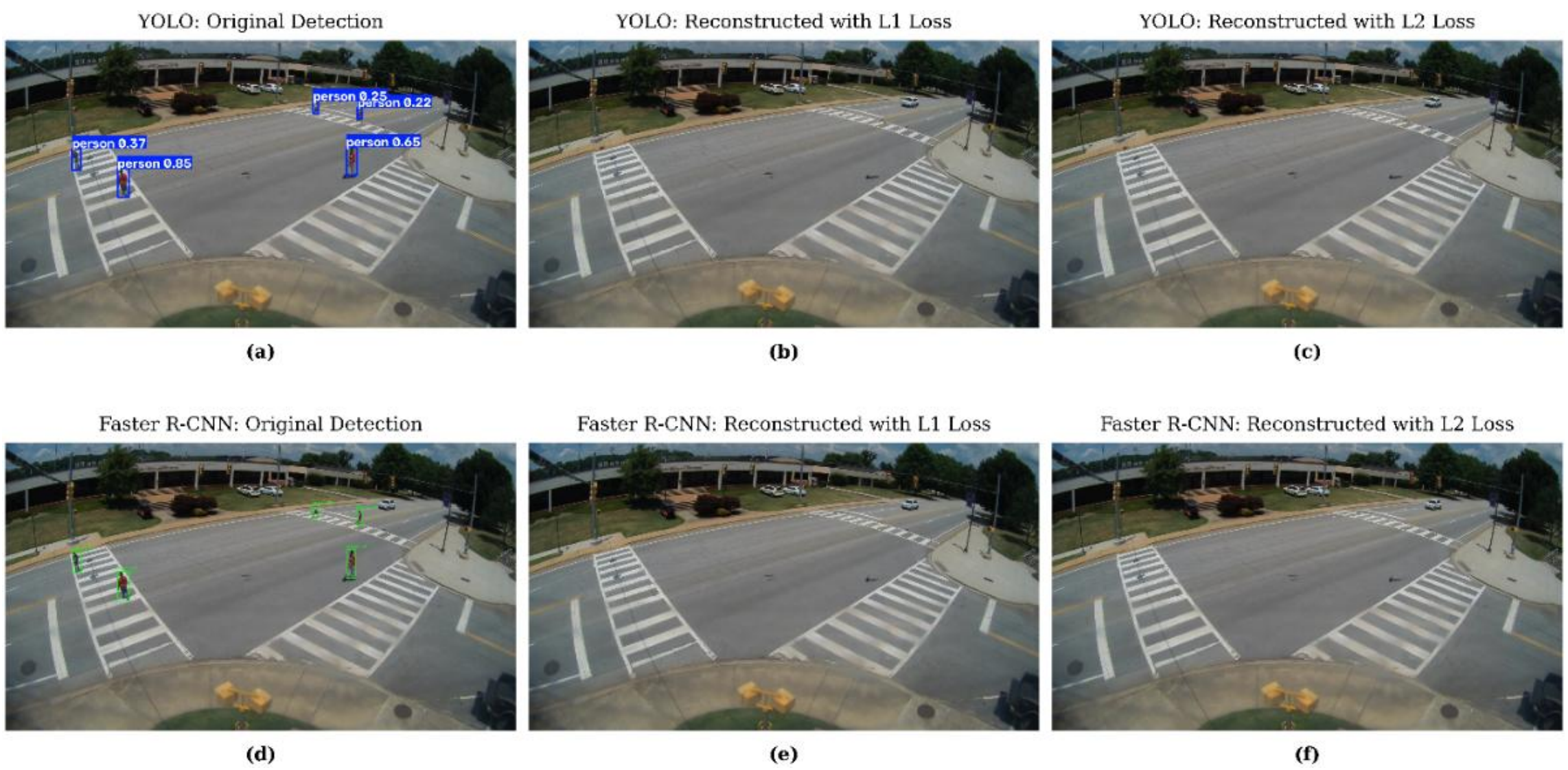

**Figure 4 Qualitative comparison of original frames captured at an intersection and reconstructed frames. Top row: (a) original frame with YOLO-based detection, (b) and (c) reconstructed frame using the L1 and L2 seam-loss formulation, respectively. Bottom row: (d) original frame with Faster R-CNN detection, (e) and (f) reconstructed frame using the L1 and L2 seam-loss formulation**

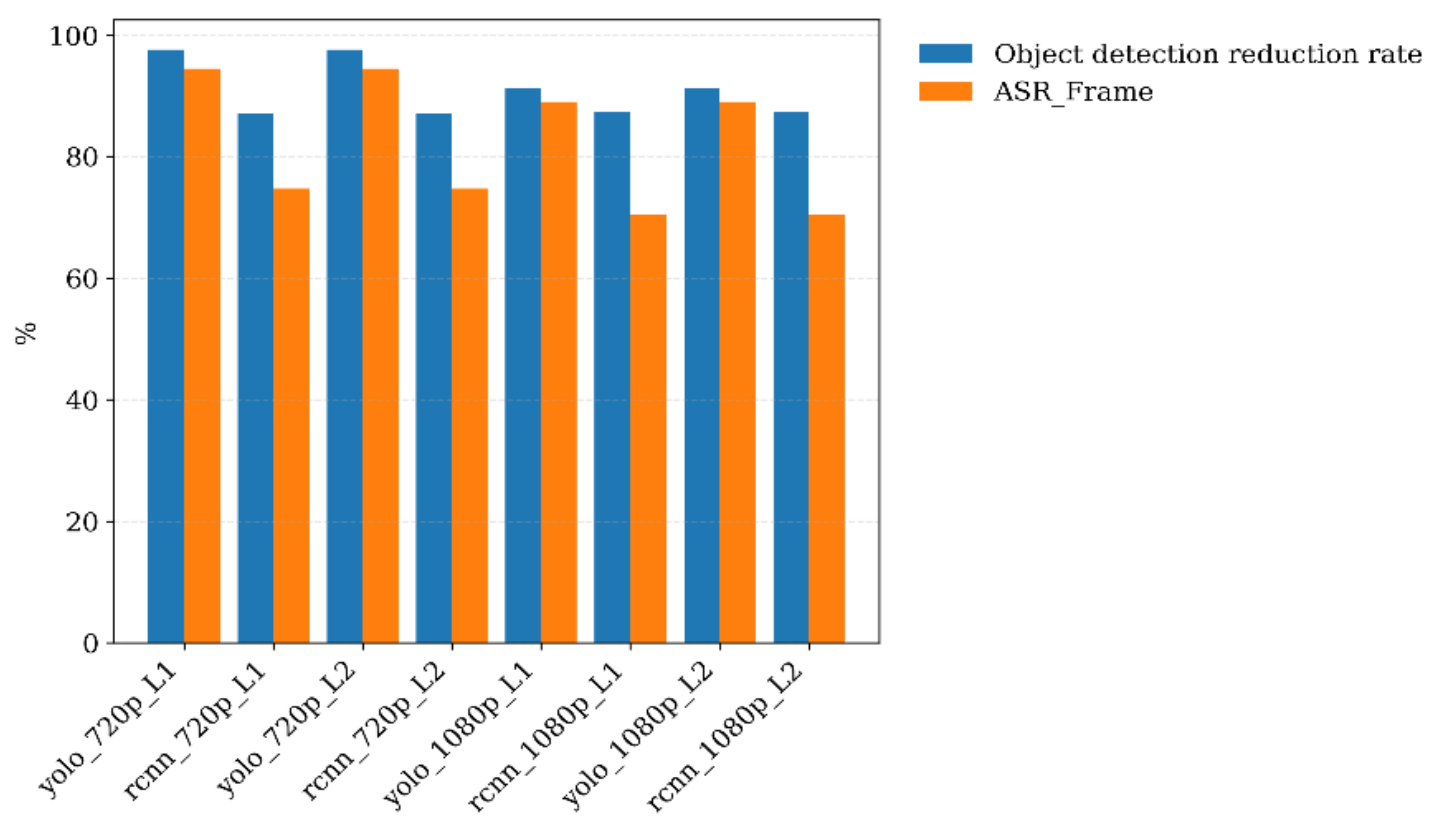

**Figure 5 Object detection reduction rate and $ASR_{frame}$ score for different object detection models, frame resolutions and boundary losses**

*Runtime Performance*

The timing results in **Figure 6** characterize the computational cost of the attack framework on an NVIDIA A100 GPU. Under the YOLO-based configuration with $L_1$ seam-loss, the mean pedestrian detection time ranged from 0.0362 to 0.0458 seconds per frame, with higher resolution frames taking longer times. The corresponding reconstruction stage times varied between 0.0377 and 0.0515 seconds per frame, respectively. Similar values were obtained with the quadratic loss formulation, indicating that the choice of seam-loss function had only a minor effect on runtime. The Faster R-CNN-based configuration required approximately twice the time required for the YOLO-based model for object detection, while reconstruction time increased slightly. The results indicate that the real-time execution of our attack relies largely on the efficiency of the object detection model, with little to no impact of the reconstruction process and loss configuration. Overall, the measured times during our experiments are suitable for near-real-time execution on high-performance GPU hardware.

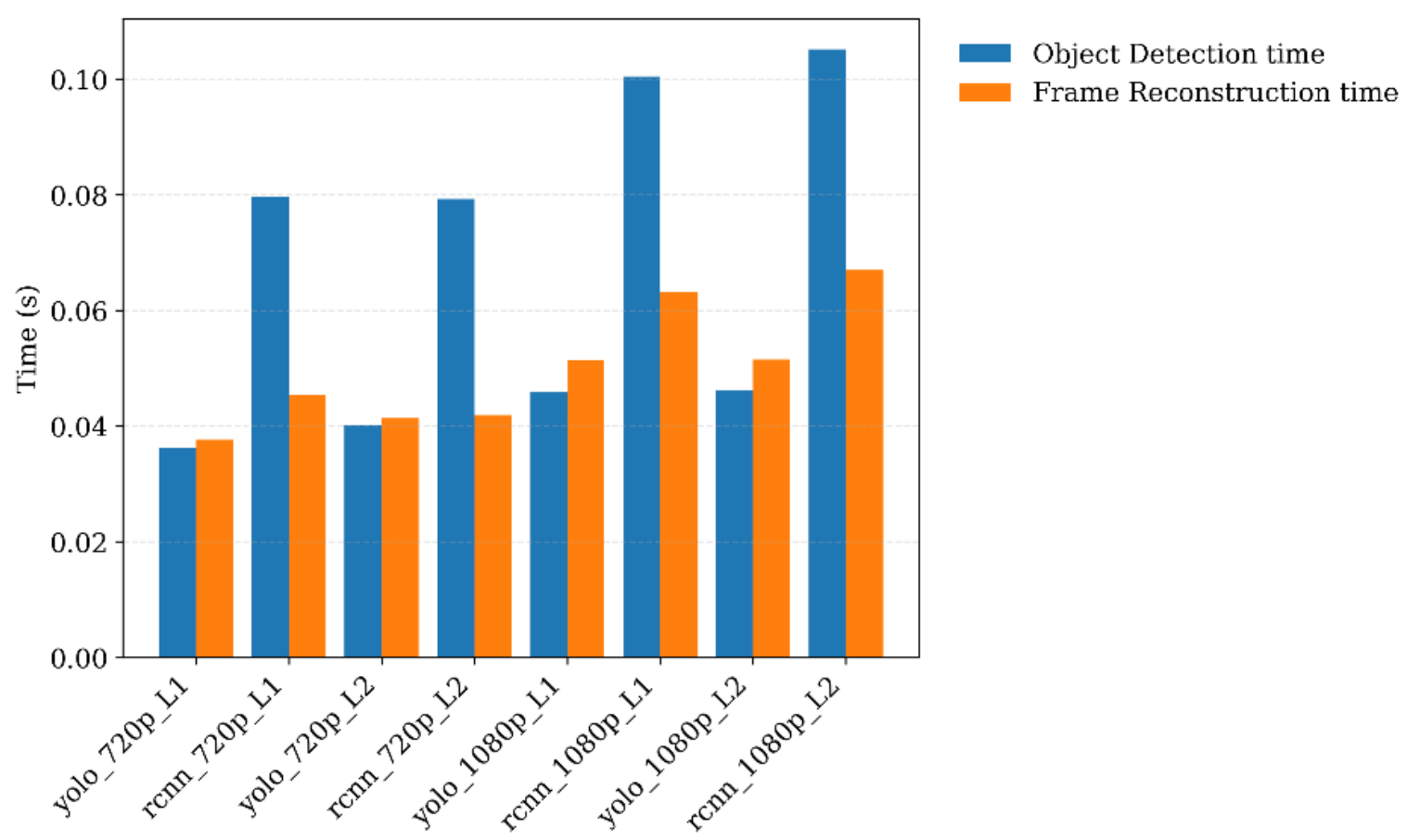

**Figure 6 Object detection time and frame reconstruction time for different object detection models, frame resolutions and boundary losses**

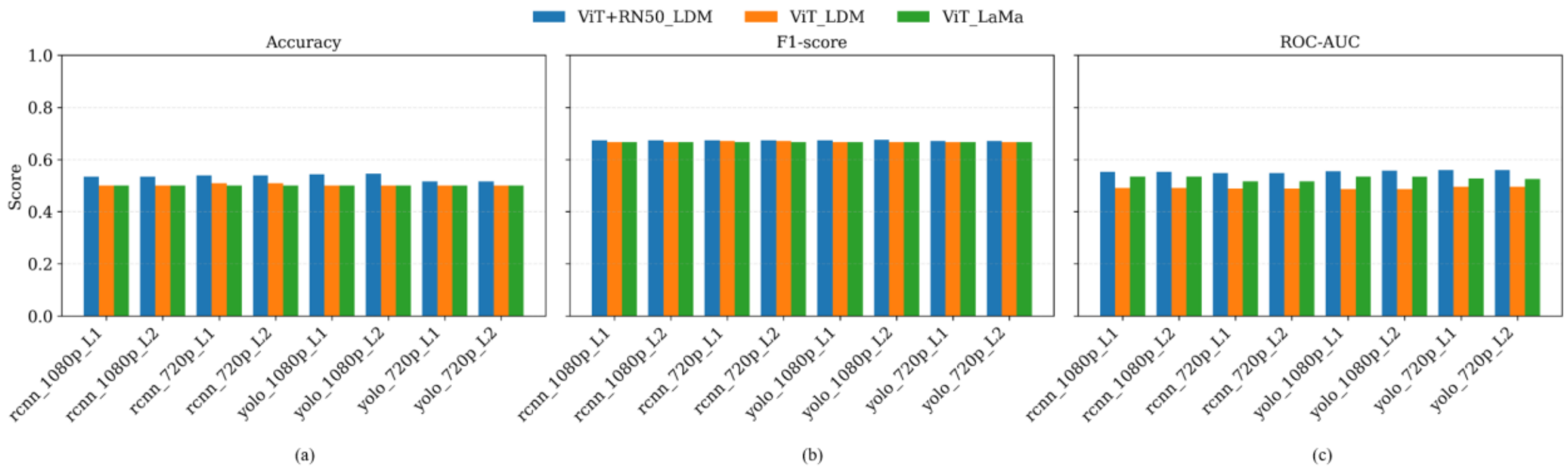

**Figure 7 (a) Accuracy, (b) F1-score, and (c) ROC-AUC value for DeCLIP model checkpoints**

*Tamper Detectability*

The tamper-detectability results are presented in **Figures 7-9**, exhibiting that performance remained similar across forensic tamper-detection models, pretrained checkpoints, and image resolutions. For the evaluated DeCLIP models, accuracy generally ranged from approximately 0.50 to 0.55, while ROC-AUC ranged from approximately 0.48 to 0.56, and F1-score remained close to 0.67 in all cases, as depicted in **Figure 7**. These values indicate limited ranking and classification capability under the tested configurations. Further analysis confirmed that the model frequently predicted class 1, thus misclassifying a lot of original frames as manipulated. Hence, the F1-score does not represent a balanced score, and these values indicate limited and inconsistent classification between real and reconstructed frames. PSCC-Net also produced results similar to the DeCLIP model, as shown in **Figure 8**. Accuracy remained at 0.50, F1-score at 0.67, and ROC-AUC ranged from 0.45 to 0.48 across the evaluated configurations, indicating that the model predominantly predicted one class. Further analysis also confirmed that the model predominantly predicted class 1. **Figure 9 (a-c)** represents the results achieved by MVSS-Net plus checkpoints on the binary classification task. Both pretrained checkpoints achieved an F1 score of approximately 0.6, an accuracy of approximately 0.50, and a ROC-AUC score ranging from 0.38 to 0.49 across the evaluated configurations. Like the previously presented models, these findings indicate inconsistent classification.
Overall, the reconstructed frames were not consistently distinguishable from authentic frames by any of the evaluated pretrained forensic models. The configurations produced performance close to random classification. The results, therefore, indicate resistance to tamper-detection under the test configurations. Models specifically trained on temporally reconstructed data may achieve better performance. However, such detection would not reverse the attack's effect on the perception system, because the target object would still remain suppressed from the attacked frame and the downstream object detector.

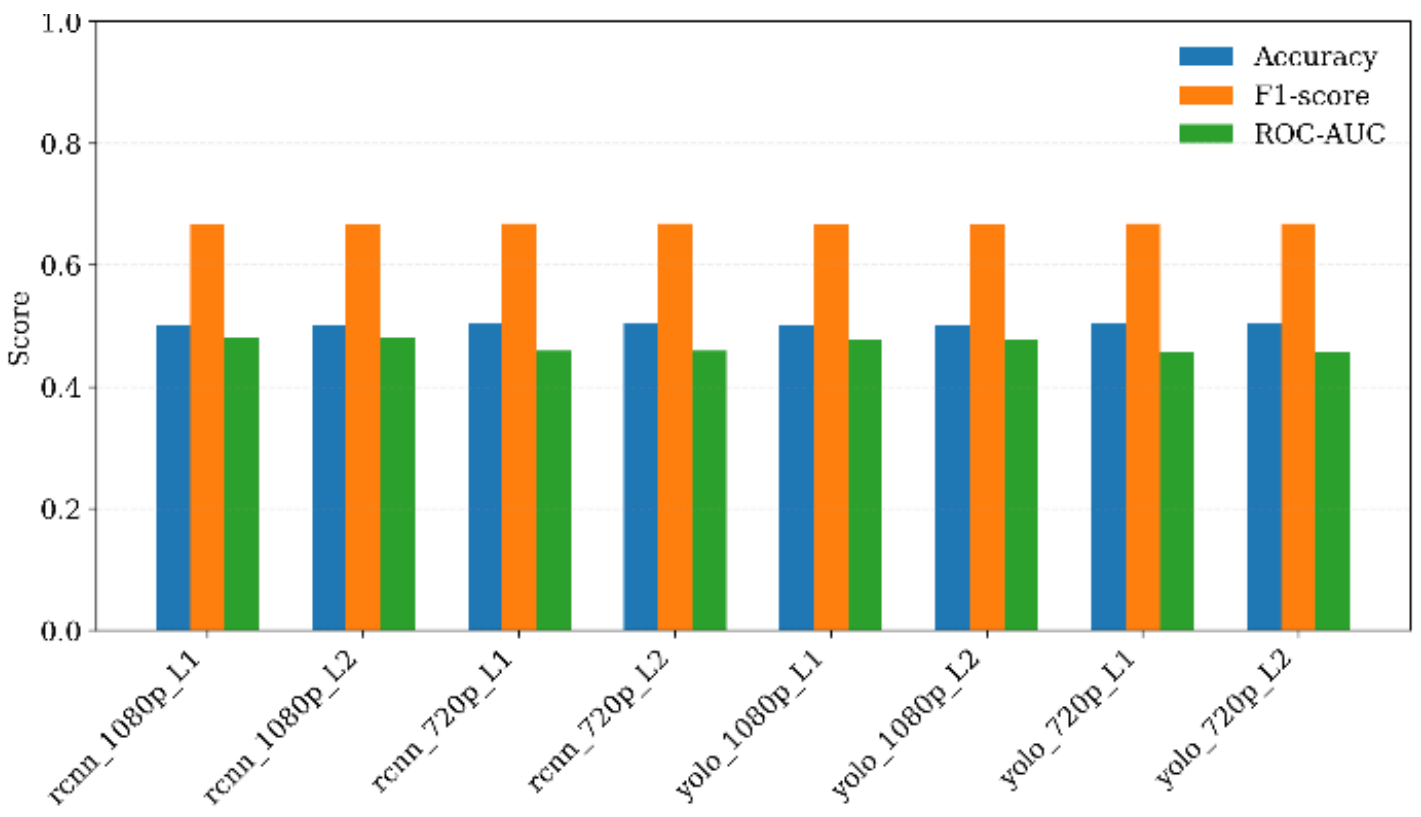

**Figure 8 Accuracy, F1-score, and ROC-AUC value for different PSCC-net model**

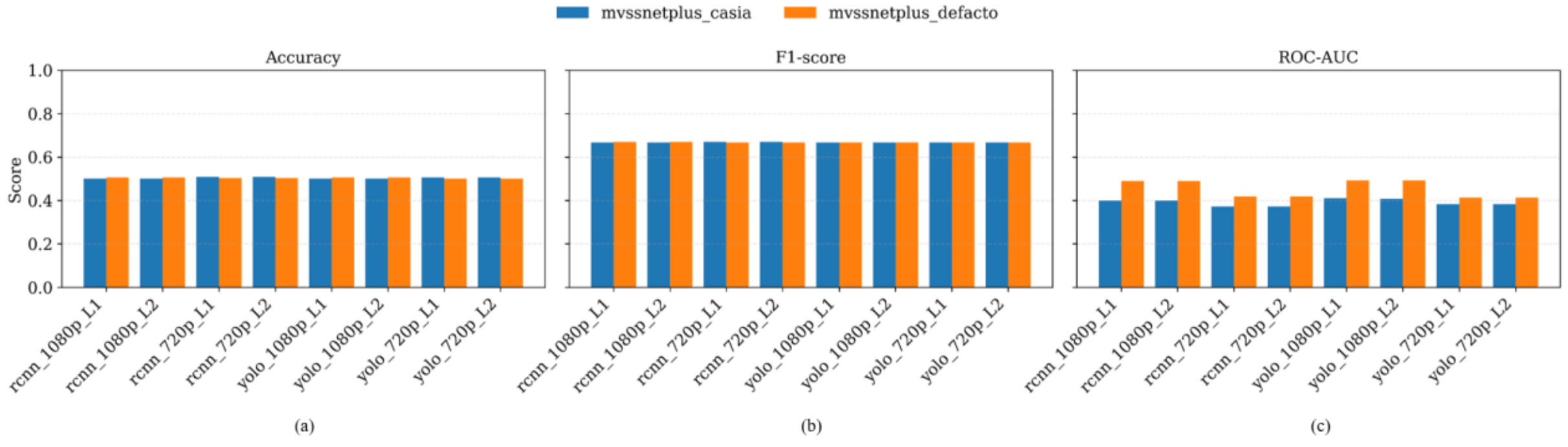

**Figure 9 (a) Accuracy, (b) F1-score, and (c) ROC-AUC score for MVSS-net plus model checkpoints**

## CONCLUSIONS

This paper presents a near-real-time targeted object-removal attack against infrastructure-mounted camera systems, demonstrating a potential threat to safety-critical intelligent transportation system (ITS) applications that rely on camera-based perception. The attack framework automatically identifies target objects, retrieves temporally consistent background content from earlier frames, and integrates the replacement using context-aware alpha blending. The attack was evaluated using video data collected from an instrumented intersection, covering multiple intersection approaches and considering two detector architectures, two image resolutions, and two seam-loss.

The experimental results show that vision-based perception can be vulnerable to temporally consistent object-removal attacks. The reconstructed frames retained high global similarity to the original inputs, while post-attack object detections by the subsequent perception system were substantially reduced. The combination of high global similarity and strong detection suppression suggests that simple visual inspection may not be sufficient to identify all manipulated frames, creating a mismatch between the actual traffic scene and the visual information available to downstream perception systems. The results obtained with both YOLO and Faster R-CNN further indicate that the attack is not limited to a single detection architecture, although its effectiveness varied with detector configuration and resolution. The forensic evaluation also showed that the pretrained tamper-detection models performed almost random classification, misclassifying frames from both authentic and tampered classes. This finding implies that existing off-the-shelf forensic models do not provide consistent protection against temporally reconstructed object-removal attacks in infrastructure-camera imagery.

Overall, the study highlights an important security risk of vision-based perception systems: visually plausible manipulation of roadside camera streams can substantially alter downstream perception models without requiring modification of the perception model itself. Such attacks, in real time or near real time, can degrade the reliability of safety-critical ITS applications, potentially leading to unsafe conditions for vulnerable road users. Beyond ITS applications, this attack could also be extended to security applications, such as airport and seaport surveillance and critical infrastructure protection, where the disappearance of a suspicious vehicle or individual from camera feeds could compromise situational awareness and security operations. These findings motivate the development of tamper-resilient roadside perception, temporal integrity verification, and cross-sensor consistency mechanisms for safety-critical applications.

Although the dataset used in the experiment covers different daytime lighting conditions, the attack could not be evaluated at night. The forensic analysis was also limited to selected publicly available pretrained models trained on datasets with characteristics similar to our attack dataset. Future work will evaluate the attack on larger and more diverse traffic datasets, including nighttime, adverse-weather, and high-density scenarios. This broader evaluation will help determine the generalizability of the attack. Future work will also investigate defenses such as cryptographic frame-integrity protection and machine learning models to detect the attack.

**ACKNOWLEDGMENTS**

The author acknowledges using large language models (LLMs), particularly OpenAI's ChatGPT, to make editorial improvements to this manuscript. LLM was solely used for refining grammar, improving readability, and formatting references.

**AUTHOR CONTRIBUTIONS**

The authors confirm contribution to the paper as follows: study conception and design: MI. Hasan, MS. Salek, M. Chowdhury, R. Ge; data collection: MI. Hasan, MS. Salek; analysis and interpretation of results: MI. Hasan, MS. Salek, N. Jones; draft manuscript preparation: MI. Hasan. All authors reviewed the results and approved the final version of the manuscript.

**DECLARATION OF CONFLICTING INTERESTS**

The authors declared no potential conflicts of interest with respect to the research, authorship, and/or publication of this article.

**FUNDING**

This work is based upon the work supported by the National Center for Transportation Cybersecurity and Resiliency (TraCR) (a U.S. Department of Transportation National University Transportation Center) headquartered at Clemson University, Clemson, SC, USA. Any opinions, findings, conclusions, and recommendations expressed in this material are those of the author(s) and do not necessarily reflect the views of TraCR, and the U.S. Government assumes no liability for the contents or use thereof.